\documentclass{cimento1}

\usepackage{graphicx,rotate}
\begin{document}

\title{Re-discovery of the top quark at the LHC and first measurements}
\author{B.~S.~Acharya\from{i1}, 
	F.~Cavallari\from{i21},
	G.~Corcella\from{i5}\from{i51},
	R.~Di Sipio\from{i4} \\ \atque 
	G.~Petrucciani\from{i51}
	}

\instlist{
	\inst{i1}  Abdus Salam International Center for Theoretical Physics and 
	 INFN Sezione di Trieste, Italy
	\inst{i21} INFN,  Sezione di Roma, Italy
	\inst{i5} Museo Storico della Fisica e Centro Studi e Ricerche E.~Fermi, Italy
	\inst{i51} Scuola Normale Superiore and INFN, Sezione di Pisa, Italy
	\inst{i4} Universit\`a di Bologna and INFN Sezione di Bologna, Italy

}

\PACSes{
\PACSit{14.65.Ha }{Top quarks} }


\maketitle

\begin{abstract}
This paper describes the top quark physics measurements that can be performed 
with the first LHC data in the ATLAS and CMS experiments.
\end{abstract}

\section{Top quark production at the LHC}
The Large Hadron Collider (LHC) is the new CERN proton-proton collider 
that will start taking data in the course of 2008. 
Initially the collisions will take place at $\sqrt{s} = 10$~TeV, moving on
to the planned $\sqrt{s} = 14$~TeV in early 2009. 

Top quarks at the LHC are produced mainly in $t\bar{t}$ pairs 
via gluon fusion or $q\bar{q}$ scattering.
The gluon-fusion mechanism accounts for about 90\% 
of the cross-section and the onset of this subprocess 
at high energy is responsible for the much larger 
cross-section at the LHC compared to the Tevatron at Fermilab. 
The total pair-production cross-section 
is estimated~\cite{top_sigma} to be $\sigma(t\bar{t})=\left(833^{+52}_{-39}\right)$~pb.
Single top production occurs via electroweak interactions,
with a total cross-section of about 300~pb.

Because $|V_{tb}|\simeq 1$, the top quark decays almost always as $t\rightarrow Wb$.
$t\bar{t}$ events can be then classified depending on how the $W$ decays as: 
{\it hadronic} if both $W$'s decay hadronically (branching ratio
BR $ \simeq ({2/3})^2 \simeq 44\%$), 
{\it semi-leptonic} if one $W$ decays leptonically and one hadronically 
(BR $\simeq 2\times 1/9 \times 2/3 \simeq 15\%$ 
for each lepton type $e$, $\mu$ or $\tau$)
or {\it di-leptonic} if both $W$'s decay leptonically
(BR $\simeq (1/3)^2\simeq 11.1\%$). 

$t\bar{t}$ events can be used to investigate many relevant observables and
properties of top quarks:  
production cross-section, top-quark mass, charge, 
polarization, rare decays, such as the ones involving flavour changing neutral 
currents (FCNC). The top quark is the heaviest and less studied quark: the 
study of its properties may open a window to new physics.
$t\bar{t}$ events are also going to be important as a `validation tool'
of many physics variables: in fact, 
$t\bar{t}$ events contain leptons, jets, $b$-jets and
missing transverse energy.
They can therefore be used to study the $b$-tagging performance
and calibrate the jet energy scale. 
The quick observation of top-quark pair events 
will give the needed credibility to possible signals 
of new physics.
Top events also constitute an important 
background for many searches beyond the Standard Model:
as a first step, the distributions of all important 
kinematic variables of these events must be well
understood. 

Most of the top properties studies are performed 
using the semi-leptonic or di-leptonic channels,
since the presence of isolated leptons and missing energy ensure 
a higher purity than the all-hadronic channel.

In this paper we describe the $t\bar{t}$ event topology, 
selection and possible measurements, 
with particular emphasis on the early data. 
The LHC luminosity will gradually increase in the first months; here
we review the experiments 
potential in the top physics sector for 
a typical integrated luminosity between 100~pb$^{-1}$ and 1~fb$^{-1}$. 

A typical selection for 100~pb$^{-1}$ has to take into account that 
the detector may not be perfectly calibrated and aligned. As a consequence,
physics analysis tools like $b$-tagging 
or missing energy calculation will not be yet fully 
commissioned.  
Trigger selections and offline selection cuts 
must then rely on simple and robust criteria. 

\section{Event selection}

\subsection{Semi-leptonic events}

Semi-leptonic events with one electron or muon in the final state are 
characterized by 
one isolated lepton, some missing transverse energy, two heavy flavour 
jets and two light jets. 
The main backgrounds are single $W$ boson production with associated jets, 
including the heavy-flavour channel ($Wb\bar{b}$+jets), 
and QCD events with heavy flavour jets which produce 
leptons from $B$ decays or fake leptons. 

In the early data-taking period, 
the most reliable triggers to study these events
will likely be the single-lepton triggers,
which typically have a threshold at $p_T\simeq 20$~GeV.
Additional cuts will then be performed offline to improve the 
lepton identification and the isolation.
At least four jets are required in the event within the tracker acceptance.
$b$-tagging can be applied to these if a particular analysis requires it.
Often a loose cut on the missing $E_T$ is applied, as well as
additional kinematic cuts on the reconstructed $W$ and top masses, 
or on the scalar sum of the transverse energies of leptons and jets.
A typical efficiency in this channel is around 5\%
and the signal over background ratio $S/B\simeq$3.

\subsection{Di-leptonic events}

Di-leptonic events with electrons or muons in the final state are characterized by 
two isolated leptons, missing transverse energy and two heavy flavour jets. 
The main backgrounds in this channel 
are single 
$Z\rightarrow l^+ l^-$ or $W\rightarrow l \nu$  production with associated jets, 
including heavy-flavoured jets.
In the $W$ boson events 
the second lepton can arise 
from the decay of heavy flavour jets or from a fake. 
Boson pair production is also an important background.

The presence of two isolated leptons allows one to select these events with high purity,
especially the $e-\mu$ channel. 
These events often fire the single- and double-lepton triggers, following which
additional cuts can be applied offline to improve the 
lepton identification and isolation.
Two jets are required in the event within the tracker acceptance 
in order to perform the $b$-tagging, if needed. A missing-$E_T$ cut
and cuts on the reconstructed dilepton mass further suppress the QCD and $Z$ backgrounds.
A typical efficiency in this channel is around 3-4\%, with a signal 
over background ratio $S/B\simeq 5-10$.

\subsection{Startup scenario}

\begin{figure}[hp]
\begin{center}
    \includegraphics[width=0.7\textwidth,angle=-90]{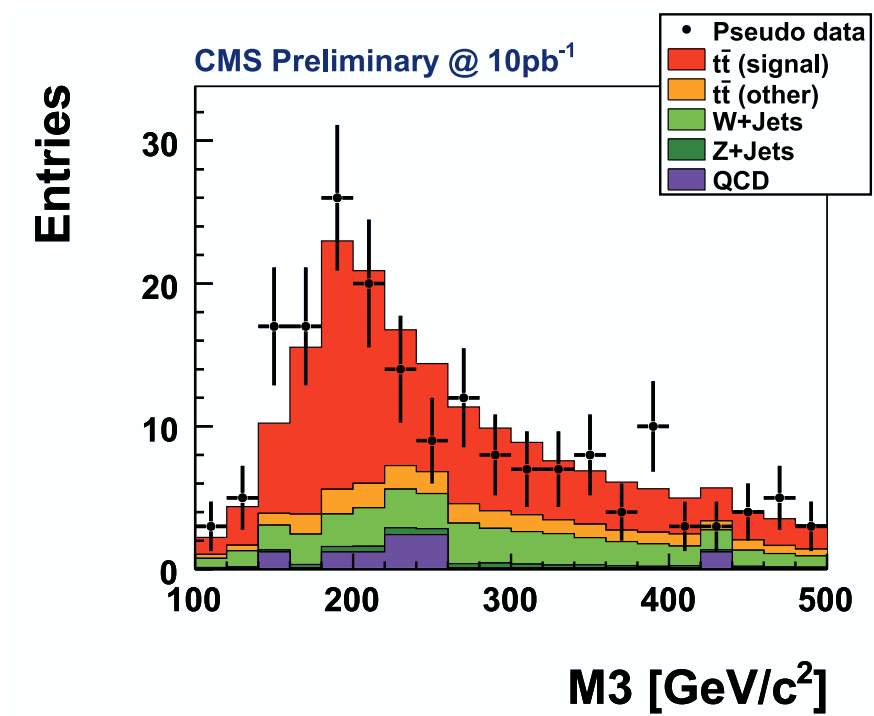}
  \caption{\small{CMS Monte Carlo study of the semileptonic muon $t\bar{t}$ 
channel with the first 
10~pb$^{-1}$ data: invariant mass of the three jets 
with the highest vectorially-summed $E_{T}$.}}
  \label{fig:masspeak_10pb}
\end{center}
\end{figure}

A CMS study~\cite{cms_10_1,cms_10_2} has shown that already with 10~pb$^{-1}$ 
it is possible to select some top-pair events and reconstruct a mass peak, as shown in 
Figure~\ref{fig:masspeak_10pb}. 
This Monte Carlo study simulates 
mis-calibrated calorimeters and a mis-aligned tracker 
in realistic startup conditions. 
Relaxed cuts are applied 
in the event selection and redundancy in the reconstruction process 
is used. For example, jets are reconstructed 
both with calorimeters and by using tracks,
and lepton-based $b$-tagging is employed, instead of methods based on
vertex reconstruction.

\section{Analysis for $\mathcal{L}$ between 100~pb$^{-1}$ and 1~fb$^{-1}$}

Both CMS and ATLAS have performed detailed Monte Carlo 
studies of the top physics reach with an integrated luminosity
between 100~pb$^{-1}$ and 1~fb$^{-1}$ \cite{cms_100,atl_100}.
A typical selection for 100~pb$^{-1}$ 
has to take into account reduced detector calibration and alignement, 
and the possibility to evaluate efficiencies
directly from the data. 

In the CMS study \cite{cms_100} the 
calorimeter calibration and tracker alignement are mis-calibrated at a level 
that is expected to be reachable with 100~pb$^{-1}$ integrated luminosity.
Jets are reconstructed using the iterative cone algorithm;
the missing transverse energy is calculated using calorimetric 
towers and taking into account the muon momentum. 
Both options to use or not the $b$-tagging are considered. In the first case, 
$b$-jets are identified using the track-counting 
algorithm and a loose $b$-tagging is generally employed. 
This simple algorithm considers the transverse impact parameter of the $N$-th track in a 
jet, where the tracks are sorted by impact parameter.
In the di-leptonic events with taus ($e\tau$ and $\mu\tau$),
the $\tau$-jet tagging algorithm is used to select a narrow, isolated calorimetric jet, 
well matched 
with one or three tracks and possibly with some indication of large impact parameter.
Techniques to evaluate the efficiency and misidentification rate for the $b$-tagging and
lepton-selection directly from the data have been studied.

In ATLAS, the cross-section is determined both in the semi-leptonic and dilepton channel. 
For the semi-leptonic channel the measurement is performed with and without relying 
on $b$-tagging. 
A strategy is studied to evaluate the $W+$jets 
background normalization using the inclusive $W$, $Z$   
and $Z+$jets cross-sections via the following relation: 
$\sigma{(W_{\mathrm{incl}})}/\sigma{(W+nj)}=\sigma{(Z_{\mathrm{incl}})}/\sigma{(Z+nj)}$, 
which is valid to a few percent level as suggested by~\cite{berends}.  

A crucial point in these analyses is the control of the QCD background, 
characterized by a very large cross-section and
small efficiency. 
Thus, extremely large Monte Carlo samples 
would be needed to correctly estimate the contamination by these processes. 
To calculate the efficiency, two techniques were used 
by the CDF and D0 experiments at the Tevatron: 
the cut factorization 
and the `matrix' method. 
The cut-factorization technique consists in finding two sets of uncorrelated cuts, 
measuring the two efficiencies separately on Monte Carlo events 
and assuming that the total efficiency is the product of the two. 
The matrix method estimates the QCD background
directly from the data by extracting the 
QCD efficiency in a signal-free control region, by loosening one of the analysis cuts.

\section{Top Physics Reach at Start-up}

Both ATLAS and CMS are putting great effort into studies 
with very low integrated luminosity. 
The first measurement is the rediscovery of the top quark:
event selection in the di-lepton and semi-leptonic channels and mass peak reconstruction.
This is within reach already with about 10~pb$^{-1}$.

With about 100~pb$^{-1}$, the $t\bar{t}$ cross-section 
can be measured with an error of the order of 15\%, 
and a first mass measurement can be given with a few hundred pb$^{-1}$. 
However, a precise mass measurement (below 1~GeV precision) requires 
detailed understanding of many systematic effects, due to both the detector 
and the techniques of analysis: 
calibrations, energy scales, missing-energy evaluation, possible 
biases in kinematic reconstructions, etc. 

\subsection{$t\bar{t}$ Invariant Mass Spectrum}
\begin{figure}[hp]
\begin{center}
    \includegraphics[width=0.7\textwidth]{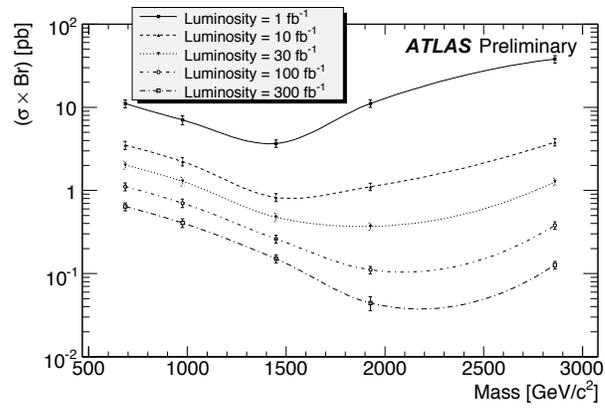}
  \caption[$t\bar{t}$ resonances]{\small{Reach for a generic
narrow resonance decaying into $t\bar{t}$ pairs in ATLAS.}}
  \label{fig:ttres}
\end{center}
\end{figure}

Starting from $\mathcal{L}=1~$fb$^{-1}$ ATLAS~\cite{atl_top_prop} 
and CMS will give a measurement of the $t\bar{t}$ 
invariant-mass spectrum. This spectrum will be sensitive to $t\bar{t}$ resonances and other
new physics effects. Figure~\ref{fig:ttres} shows the expected 
ATLAS reach for a generic narrow $t\bar{t}$ mass resonance.

\subsection{Single Top Events}
Besides $t\bar t$-pair production,
single top quarks will also be produced via
electroweak processes at the LHC. 
There are three main production channels:
the $t$-channel has the largest cross-section ($\sim 240$~pb),
followed by $tW$ associated production ($\sim 60$~pb) and finally the $s$-channel 
($\sim 10$~pb).
The single-top 
production cross-section can give a measurement of the CKM matrix element V$_{tb}$.
The most important backgrounds come from  $t\bar{t}$ and $W$+jets events, 
which are difficult to reduce.
An analysis by ATLAS~\cite{atl_singletop} has shown statistical significance already for 
$\mathcal{O}(100~$pb$^{-1})$. 
A study by CMS in~\cite{cms_tdr} shows that the $t$-channel will be easily observable
with 10 fb$^{-1}$~\cite{cms_tdr}. 
Furthermore, rescaling the statistical and systematic uncertainties 
obtained with the 10~fb$^{-1}$ analysis, it is found that    
single-top production in the $t$-channel 
can also be visible at $5\sigma$ confidence level even with a luminosity of 1~fb$^{-1}$.

\subsection{Top Charge}
A preliminary study by ATLAS~\cite{atl_top_prop} has concluded that 
it is possibile to measure the top 
charge by pairing the lepton and the $b$-jet via their invariant mass. 
The jet-charge computation is made 
from the charges of the jet tracks weighted 
by their proximity to the jet axis. 
A discrimination between charges $+2/3$ and $-4/3$ is 
possible with 1~fb$^{-1}$.

\subsection{FCNC in Top Decays}
Searches for flavour changing neutral 
currents in top-quark decays will benefit from the high
luminosity and large production cross-section for $t\bar{t}$ pairs.
According to a study by CMS~\cite{cms_tdr},
with 10~fb$^{-1}$ $t\to qZ$ and $t\to q\gamma$ decays will be
observable with $5\sigma$ significance down to a BR $\sim 10^{-3}$,
comparable to the estimated signals in some Beyond-SM models, such as
$R$-parity violating SUSY.
Preliminary results \cite{atl_top_prop} 
show that even with an integrated luminosity of 1~fb$^{-1}$ 
ATLAS is sensitive to FCNC decays with BR$(t\to q Z) \sim
10^{-3}$ and BR$(t\to q\gamma) \sim 10^{-4}$.

\section{Conclusions}
The LHC will give high statistics samples of $t\bar{t}$ pairs
that will allow detailed studies of the top quark properties. 
Rediscovery of the top quark is possible already at 
very small integrated luminosity (10~pb$^{-1}$).
Such a quick confirmation of top-quark pair events 
will give credibility to 
possible signals of new physics in the ATLAS and CMS experiments.
With more luminosity, ATLAS and CMS will measure precisely many important properties of top quarks and
will be sensitive to various 
sources of new physics, such as $t\bar{t}$ resonances and flavour changing neutral
currents. The importance of top quark events at the LHC cannot be emphasised enough.

%


\begin{thebibliography}{0}

\bibitem{top_sigma}
{\em  NLL resummation of the heavy-quark hadroproduction cross section,}
\BY{Roberto Bonciani, Stefano Catani, Michelangelo Mangano}
\IN{Nucl Phys B}{ 529}{1998}{ 424-450}
\bibitem{cms_10_1} 
{\em Expectations for observation of top quark pair
production in the dilepton final state with the 
first $10~$pb$^{-1}$ of CMS data},
\BY{The CMS Collaboration}
  {CMS PAS TOP-08-001}{2008}
\bibitem{cms_10_2} 
{\em Observability of 
Top Quark Pair Production in the Semileptonic 
Muon Channel with the first 10~pb$^{-1}$ of CMS Data}
\BY{The CMS Collaboration}
  {CMS PAS TOP-08-005}{2008}
\bibitem{cms_100} {\em Measurements of the $t\bar{t}$ cross section 
in the dilepton channel with 
 {$\mathcal{L}=100$~pb$^{-1}$} using the CMS Detector,}
\BY{The CMS Collaboration}
  {CMS PAS TOP-08-002}{2008}
\bibitem{atl_100} {\em Determination of the Top quark production cross-section in ATLAS}
\BY{The ATLAS Collaboration}
{ATLAS Note in preparation}.
\bibitem{berends}
\BY{F.A.Berends et al.}
\IN{Phys. Lett. B}{224}{1989}{237}
\bibitem{atl_top_prop}{\em Top Quark Properties}, 
\BY{The ATLAS Collaboration}
{ATLAS Note in preparation}.
\bibitem{atl_singletop} {\em Prospects for Single top measurements in ATLAS}
\BY{The ATLAS Collaboration} {ATLAS Note, in preparation.}
\bibitem{cms_tdr}
{CMS Physics TDR: Volume II (PTDR2), Physics Performance: 
CERN-LHCC-2006-021, 25 June 2006.} 


\end{thebibliography}
\end{document}